\documentclass[aps,twocolumn,prl,amsmath,showpacs,superscriptaddress,floatfix]{revtex4}

\usepackage{graphicx}
\usepackage{dcolumn}
\usepackage{bm}

\begin{document}

\title{Interfaces in driven Ising models: shear enhances confinement}

\author{Thomas H. R. Smith}
\affiliation{H.H. Wills Physics Laboratory, University of Bristol,
  Tyndall Avenue, Bristol BS8 1TL, United Kingdom}
\author{Oleg Vasilyev}
\affiliation{Max-Planck-Institut f{\"u}r Metallforschung,
  Heisenbergstra\ss e~3, D-70569 Stuttgart, Germany} 
\author{Douglas~B.~Abraham}
\affiliation{Max-Planck-Institut f{\"u}r Metallforschung,
  Heisenbergstra\ss e~3, D-70569 Stuttgart, Germany} 
\affiliation{Theoretical Physics, Department of Physics, University of
  Oxford, 1 Keble Road, Oxford OX1 3NP, United Kingdom} 
\author{Anna Macio\l ek}
\affiliation{Max-Planck-Institut f{\"u}r Metallforschung,
  Heisenbergstra\ss e~3, D-70569 Stuttgart, Germany} 
\affiliation{Theoretical Physics, Department of Physics, University of
  Oxford, 1 Keble Road, Oxford OX1 3NP, United Kingdom} 
\affiliation{Institute of Physical Chemistry, Polish Academy of Sciences,
Department III, Kasprzaka 44/52, PL-01-224 Warsaw, Poland}
\author{Matthias Schmidt}
\affiliation{H.H. Wills Physics Laboratory, University of Bristol,
  Tyndall Avenue, Bristol BS8 1TL, United Kingdom}
\date{12 June 2008}

\begin{abstract}
  We use a phase-separated driven two-dimensional Ising lattice gas to
  study fluid interfaces exposed to shear flow parallel to the
  interface.  The interface is stabilized by two parallel walls with
  opposing surface fields and a driving field parallel to the walls is
  applied which (i) either acts locally at the walls or (ii) varies
  linearly with distance across the strip. Using computer simulations
  with Kawasaki dynamics, we find that the system reaches a steady
  state in which the magnetization profile is the same as that in
  equilibrium, but with a rescaled length implying a reduction of the
  interfacial width. An analogous effect was recently observed in
  sheared phase-separated colloidal dispersions. Pair correlation
  functions along the interface decay more rapidly with distance under
  drive than in equilibrium and for cases of weak drive can be
  rescaled to the equilibrium result.
\end{abstract}


\maketitle


Shearing a fluid is a fundamental example in the statistical mechanics
of systems driven away from equilibrium \cite{evans04}.  Profound
effects on the liquid structure \cite{recentShearStructure} and new
phase transitions \cite{onuki97} can result.  In a recent intriguing
advance in colloidal science interfacial fluctuations of capillary
wave-type were observed by direct optical means \cite{aarts04science};
these fluctuations can be strongly supressed by shearing in the
direction parallel to the interface~\cite{derks06prl}.

Kinetic Ising models which have been driven out of equilibrium are of
continuing importance in condensed matter physics
\cite{schmittmann,leung,cirillo05}. Underlying is the idea of coarse
graining in time \cite{penrose}, where a record of the dynamical state
is only kept at discrete points in time, which is an approximation
that can often be justified by the quality and range of deductions
made. Mass is conserved, but there is no inertia and associated
hydrodynamics. Nevertheless currents are predicted as well as
densities. Examples showing the value of such stochastic models
include studies of interfaces under spatially homogeneous and random
driving fields \cite{leung}, domain growth \cite{cirillo05} and
nucleation \cite{allen08} under shear, as well as of the precursor
film in complete wetting~\cite{dba1}.

In this Letter, we investigate the effects of shear on the interface
between coexisting phases using the two-dimensional ($2d$) Ising
lattice gas with Kawasaki dynamics \cite{kawasaki}.  The interface is
induced by fixing spins on the boundaries of a strip; the analogue of
shearing in a lattice gas will be explained below.  Studying the $2d$
problem is interesting in its own right, because in equilibrium the
interface is known to be rough for all temperatures $T$ below the bulk
critical temperature $T_c$.  If the interface is confined to a strip
of width $L_y$, then it sweeps out essentially the entire strip,
meandering back and forth between the walls, such that the interfacial
width $w\sim L_y$, despite the entropic repulsion from the walls
\cite{maciolek}.  Consequently, at fixed $T$, the magnetization
profile as a function of distance $y$ across the strip scales as
$m_{\rm b}(T) {\cal M}_{\rm eq}(y/L_y)$, where $m_{\rm b}(T)$ is the
spontaneous magnetization in bulk and ${\cal M}_{\rm eq}$ is the
finite size scaling function for a strip with size $L_x$ ($\gg L_y$)
along the walls; ``eq'' labels equilibrium quantities.

Using Monte Carlo (MC) simulations we find that the $2d$ Ising model
under shear-like drive parallel to the interface achieves a steady
state that is characterized by an {\em effective} length scale
$L_y^\star$, such that the magnetization profile $m(y)$ at given $T$
obeys
\begin{equation}
\label{eq:scaling}
\frac{m(y)}{m_{\rm b}(T)}\approx
 {\cal M}_{\rm eq}\left(
 \frac{y}{L^{\star}_y}\right)
 + {\cal M}_{\rm corr}(y) \quad {\rm with} \quad L^{\star}_y < L_y,
\end{equation}
where ${\cal M}_{\rm corr}(y)$ is a boundary correction term, which
decays away from the boundaries on the scale of the bulk correlation
length~$\xi$ (on which the exponential decay of spin-spin correlations
occurs in the homogeneous system).  A conjecture based on
Eq.~(\ref{eq:scaling}) is that shear-like drive acts as effective
confinement of the system.

\begin{figure}
\includegraphics[width=0.4\textwidth]{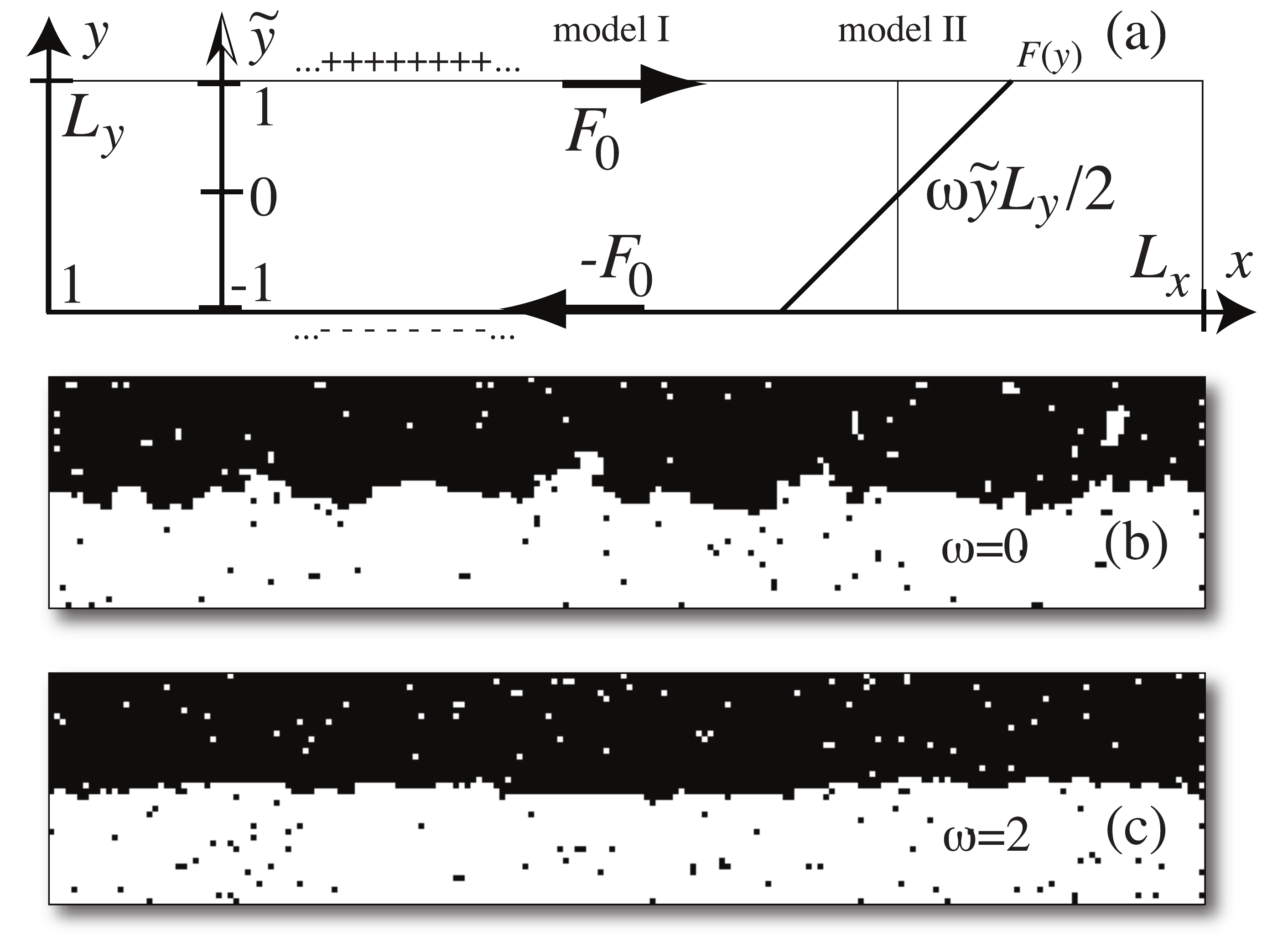}\\
\vspace{-2mm}
\includegraphics[width=0.5\textwidth]{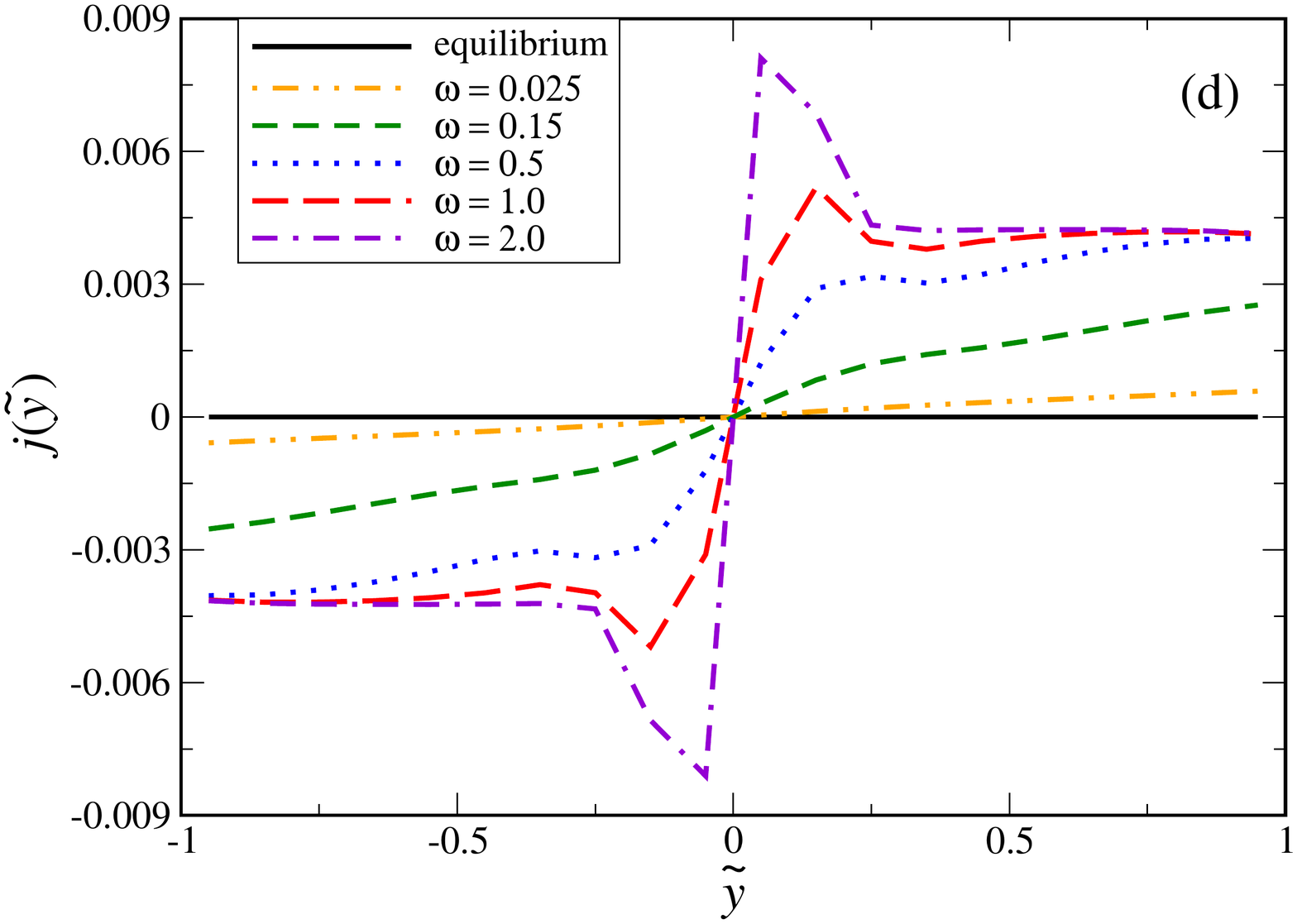}
\caption{(color online) a) Illustration of the $2d$ Ising strip
  between $+/-$ walls under an external driving field $JF(y)$ parallel
  to the walls. The field acts with strength $F_0$ at the boundary
  layers only (model I) or varies linearly with (scaled) distance
  $\tilde y$ across the strip, where $\omega$ determines the slope
  (model II).  b) Snapshot from simulation for $T/T_c=0.75, L_x=200$,
  and $L_y=20$ at equilibrium; black (white) regions indicate
  $\sigma_i=+1 (-1)$. c) Same as b) but under strong drive according
  to model II with $\omega=2.$ d) The current profile $j(\tilde y)$
  parallel to the walls as a function of $\tilde y$ for different
  values of $\omega$ (as indicated) in model II for the same values of
  $T,L_x,L_y$ as in~b.
\label{fig1}}
\end{figure}

To be specific, in spin language the Hamiltonian is
$H=-J\sum_{<i,j>}\sigma_i\sigma_j$, where the sum runs over nearest
neighbour sites $i,j$; $J>0$ is the spin-spin coupling constant, and
the spins take on values $\sigma_i=\pm 1$, corresponding to particle
occupation numbers $\tau_i=(\sigma_i+1)/2=0,1$. The interface is
induced and localized by two walls of fixed spins $\sigma_i=+1$ at the
top ($y=L_y+1$) and $\sigma_i=-1$ at the bottom ($y=0$) edges of the
strip. Periodic boundary conditions are applied in the $x$-direction.
In order to induce shear, we apply a force field $JF(y)$ parallel to
the $x$-direction.  In model I only the particles in the layer
adjacent to each wall are driven along the walls and in opposite
directions: $F(L_y)=F_0$ and $F(1)=-F_0$, and $F(y)=0$ otherwise; for
$F_0\to\infty$ these layers form an asymmetric exclusion processes
\cite{schmittmann}, coupled to an Ising strip.  In model II the force
varies linearly across the strip, such that it vanishes in the middle
of the slit: $F(y)=\omega[y-(L_y+1)/2]$, where $\omega = \partial
F(y)/\partial y$ is the (dimensionless) field difference between
adjacent rows; we also use a scaled variable $\tilde
y=(2y-L_y-1)/L_y$, see Fig.~\ref{fig1}a.  Model~II can be thought of
as mimicking the effects caused by the flow of a background solvent
\cite{derks06prl}.  The system evolves under spin-exchange Kawasaki
dynamics \cite{kawasaki}, corresponding to particle hopping to an
(empty) nearest neighbour site. The work done by (or against) the
force field, $\Delta F$, in a trial move enters a modified Metropolis
acceptance rate, $\min\{1,\exp(-(\Delta H+\Delta F)/(k_BT))\}$, where
$\Delta H$ is the change in internal energy and $k_B$ is the Boltzmann
constant.  The dynamics capture the local conservation of particle
number and the competition of forced transport with diffusive motion.

We have performed extensive simulations using single-spin and
multi-spin coding techniques \cite{gemmert05}, the latter generalized
to include drive; this facilitates simultaneously running 64 systems
per processor core. We present here results for $L_x=200$ and $L_y$
varying from 10 to 40 at fixed total magnetization
$m=\sum_i\sigma_i=0$ for $T/T_c=0.75, 0.85$, and $0.95$ (where
$T_c=2.2619J/k_B$), so that $L_y/\xi>1$. Sampling large-scale
interfacial fluctuations with inherently slow Kawasaki dynamics is
challenging and requires run lengths of the order of $N_{\rm MC}=10^8$
MC steps ($L_x\times L_y$ trial moves form one MC step).  For each run
we perform initially $N_{\rm MC}$ MC steps and find that thereafter
all observables of interest fluctuate around their mean value. We
conclude that a steady state is reached, in which we gather statistics
for a further $N_{\rm MC}$ steps. Statistical errors are obtained with
blocking and bootstrap techniques.  Fig.~\ref{fig1} displays snapshots
of configurations~\cite{movies}.

Driving creates a non-vanishing current profile $j(y)$ parallel to the
walls; $j(y)$ gives the net (mean) number of particles that move from
$x$ to $x+1$ at given $y$ in one MC step. For the boundary-driven case
(model I) the current is localized at both walls, $y=1, L_y$, and
vanishes in the middle of the system (apart from a very small
anti-current at $y=2,L_y-1$). In the bulk-sheared case (model II)
$j(\tilde y)$ varies smoothly with $\tilde y$, see
Fig.~\ref{fig1}d. For small values of $\omega$ a near-linear behaviour
is observed. For strong drive the current saturates upon approaching
the walls. Surprisingly $|j(\tilde y)|$ displays pronounced maxima
offset from the center of the strip, indicating an optimum between
sufficiently strong drive and sufficiently large number of
particle-hole pairs.

\begin{figure}
\includegraphics[width=0.47\textwidth]{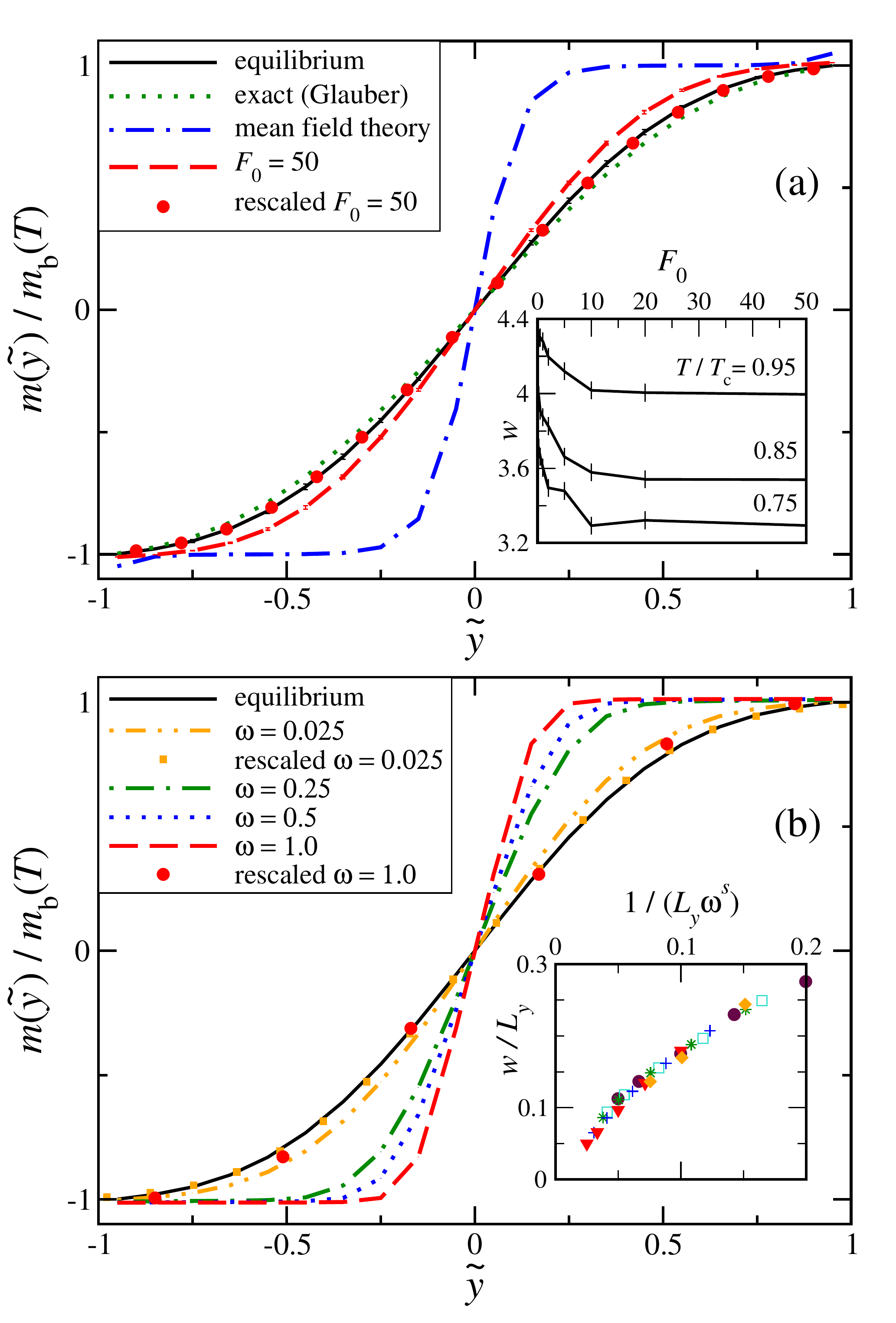}
\caption{(color online) Scaled magnetization profiles $m(\tilde
  y)/m_{\rm b}(T)$ as a function of the scaled distance~$\tilde y$ for
  $T/T_c=0.75, L_x=200$, $L_y=20$.  a) Model I: Kawasaki equilibrium
  result, Glauber equilibrium result, mean-field result, Kawasaki
  result for $F_0=50$, and the latter rescaled as $m(0.83\tilde
  y)$. Inset: Variation of the interfacial width $w$ with $F_0$ for
  three different temperatures (as indicated). b) Kawasaki simulation
  results for model~II are shown in equilibrium and for a range of
  values of the field gradient $\omega$ (as indicated), as well as
  using rescaling $m(a_\perp \tilde y)$, where $a_\perp=1/1.15, 1/3.4$
  for $\omega=0.025,1$, respectively. Error bars are of the order of
  the line thickness. Inset: Variation of the scaled interfacial width
  $w/L_y$ with the scaling variable $1/(L_y \omega^s)$, where $s=0.3$,
  for $\omega=0.025$ (diamonds), 0.1 (filled circles), 0.2 (squares),
  0.25 (stars), 0.5 (crosses), and 1 (triangles).
\label{fig2}}
\end{figure}

Having established the occurence of shear flow, we investigate the
interfacial magnetization profile
$m(y)=L_x^{-1}\langle\sum_x\sigma(x,y)\rangle$, where the angles
denote an average in the steady state. We have checked that the aspect
ratio $L_x/L_y$ is large enough such that at equilibrium the exact
result \cite{maciolek} for $m(y)$, for $L_x\to\infty$, is
indistinguishable within statistical errors from our simulation
results (not displayed) using Glauber dynamics \cite{glauber}, where
$m$ can fluctuate.  Using Kawasaki dynamics (that constrain $m=0$)
leads to larger flat regions near the walls and a sharper interfacial
region, see Fig.~\ref{fig2}a; this effect increases strongly if
$L_y/\xi \gg 1$. The importance of the contribution of capillary-wave
type fluctuations to $m(y)$ can be gauged by the stark contrast to the
kink-like behaviour obtained in equilibrium mean-field theory.
For cases of weak boundary drive (small values of $F_0$) the
magnetization near the walls increases and the interfacial region gets
squeezed (becomes narrower) upon increasing $F_0$; this behaviour is
more apparent for higher values of $T$.  For $F_0\gtrsim 5$ there are
no further changes in the profile within error bars; in
Fig.~\ref{fig2}a the limiting behaviour, $F_0=50$, is shown.  The
effect on $m(y)$ arises from a competition of advection and diffusion:
the boundary drive helps to break up clusters into smaller
constituents; these possess high mobility and are hence able to
migrate towards and coalesce with their bulk phase for energetic
reasons. This dynamic mechanism makes the observed additional
confinement of the interface physically reasonable.  The effect of the
boundary drive cannot be mimicked by assuming equilibration in static
boundary fields acting at $y=2$ and $L_y-1$. Even infinitely strong
surface fields $h(y=1)=-h(y=L_y)=\infty$ induce a significantly weaker
effect.  Particularly striking is the scaling behavior of $m(y)$, see
Eq.\ (\ref{eq:scaling}), which indicates that the effect of the
boundary drive on the interfacial region of the profile is the same as
that induced by increasing the confinement, and hence reducing the
width of the strip from $L_y$ to $L^{\star}_y=0.83L_y$ for $F_0
\gtrsim 5$. This corresponds to the scaling $m(a_\perp \tilde
y)\approx m_{\rm eq}(\tilde y)$, where for strong boundary drive (such
that essentially $F_0\to\infty$) we find $a_\perp=L_y^\star/L_y=0.83$
for $L_x=200$ and all values of $T$ and $L_y$ considered.  We expect
that $w\sim L_y^\star$, which we have checked explicitly by
calculating $w$ as the second moment of $\partial m(y)/\partial y$,
showing that $w$ decreases rapidly with $F_0$ and eventually saturates
at a finite value, see the inset in Fig.~\ref{fig2}a. Model II
displays similar squeezing of the interfacial region, see
Fig.~\ref{fig2}b. For small values of $\omega$ the effect is very
similar to the boundary-driven case. Increasing $\omega$ leads to very
pronounced squeezing, such that the central section of the profile for
$\omega=1$ resembles the mean-field solution in equilibrium! Even this
case can be rescaled to the equilibrium profile, see Fig.~\ref{fig2}b.
We find that the variation of $w$ with transversal system size and
field strength can be condensed into dependence on the scaling
variable $1/(L_y \omega^s)$ for any value of $T$ considered; for
$T/T_c=0.75$ data collapse is achieved for $s=0.3$, see the inset of
Fig.~2b.  Similar quality of scaling can be obtained by including
$(T-T_c)/T_c$ in the scaling variable \cite{leung}; these results (not
shown) indicate that in cases of weak drive $a_\perp=L_y^\star/L_y$
increases (and hence the effective confinement becomes weaker) with
increasing $T$.

\begin{figure}
\vspace{-7mm}
\includegraphics[width=0.47\textwidth]{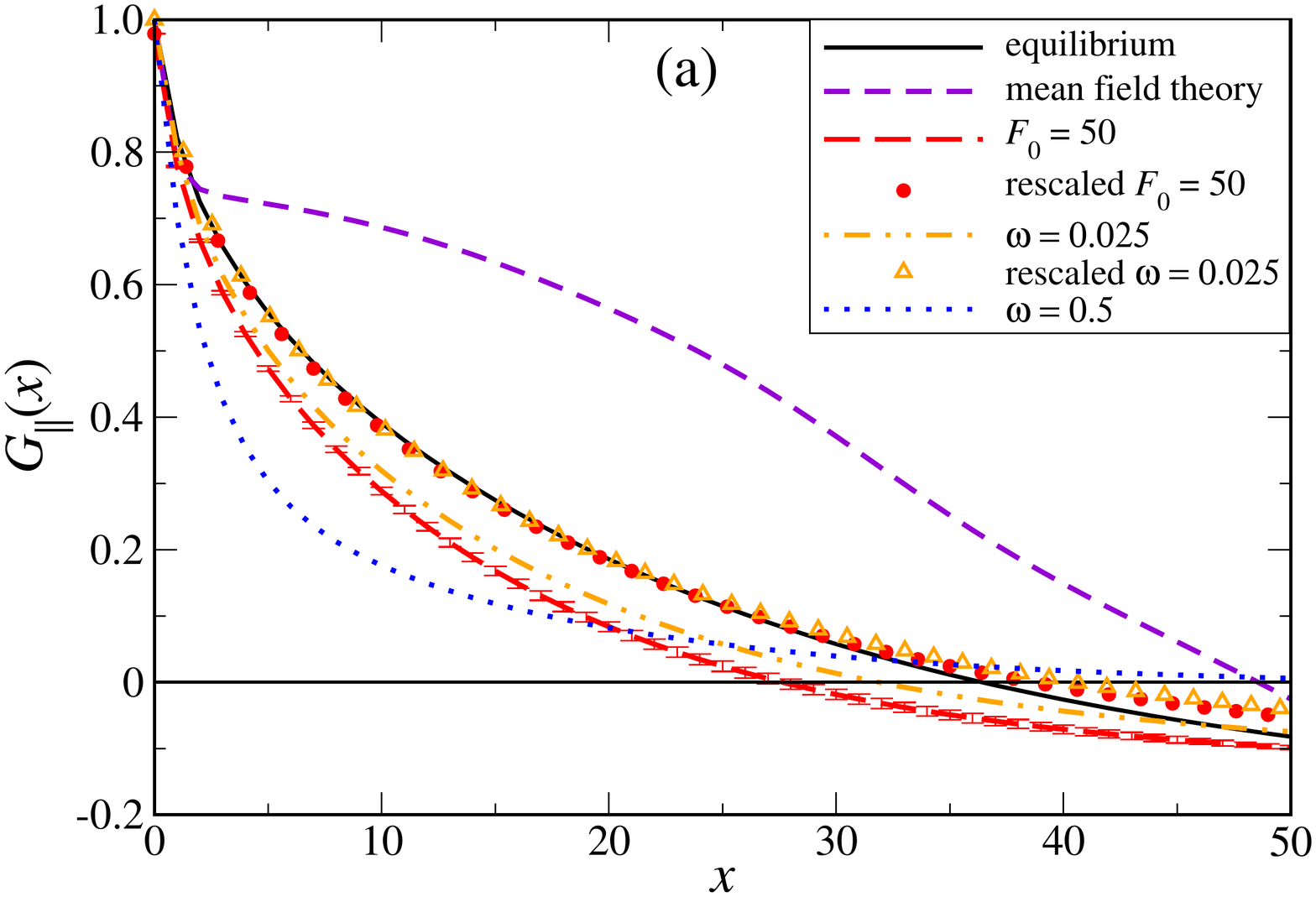}\\
\vspace{-5mm}
\includegraphics[width=0.47\textwidth]{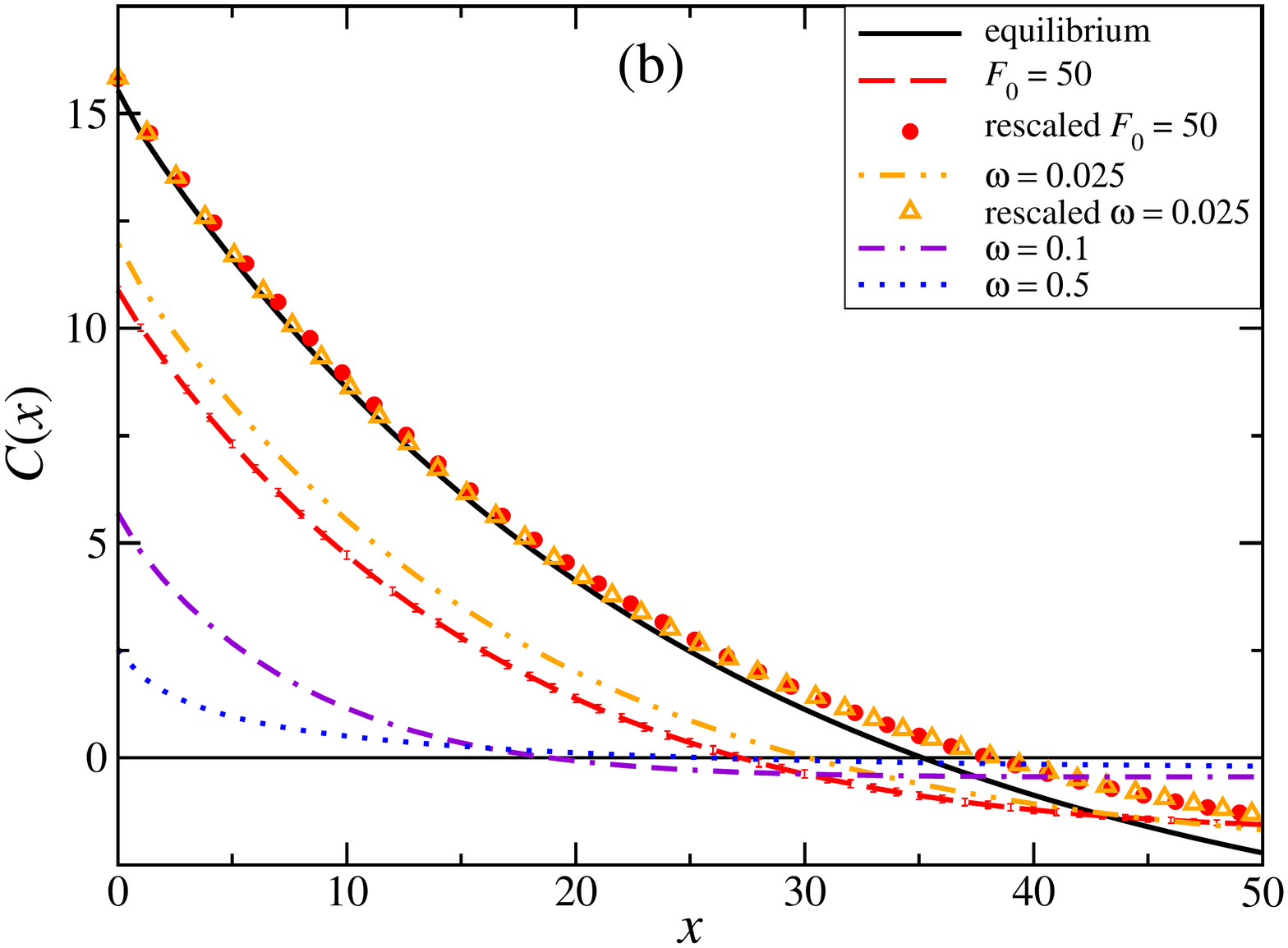}
\caption{(color online) Spin-spin pair correlation function
  $G_{\parallel}(x)$ at the center of the strip (a) and height-height
  correlation function $C(x)$ (b) as a function of distance $x$ for
  $T/T_c=0.75$, $L_x=200$, $L_y=20$.  Shown are results in
  equilibrium, from mean-field theory (only for $G_\parallel(x)$ in
  a), from Kawasaki simulations with strong boundary-drive in model~I
  ($F_0=50$), the same rescaled as $G_\parallel(x/1.4)$; Kawasaki
  simulation results for model II are shown for $\omega=0.025$, 0.1
  (only for $C(x)$ in b) and 0.5, as well as the former case rescaled
  as $G_\parallel(x/1.27)$.
}
\label{fig3}
\end{figure}

We next investigate the interfacial structure on the two-body level
and consider the spin-spin pair correlation function in steady-state
$G(x,y,y')= \langle\sigma_i\sigma_j\rangle$, where $i=(0,y)$ and
$j=(x,y')$, focusing on the behaviour along the center,
$G_\parallel(x)\equiv G(x,L_y/2,L_y/2)$; this should reveal most
clearly interface-mediated correlations. The simulation results shown
in Fig.~\ref{fig3}a indicate that shear induces a faster decay of
$G_\parallel(x)$ with distance~$x$ than at equilibrium. For model~I
and for weak drive in model~II, the behaviour at short and
intermediate distances can be accurately described by rescaling the
{\em lateral} coordinate, $G_\parallel(a_\parallel x)\approx G^{\rm
  eq}_\parallel(x)$. From fitting the data we find the ratio of the
lateral (interfacial) corrrelation lengths in and out of equilibrium,
$a_\parallel=\xi_\parallel/\xi_\parallel^{\rm eq}<1$, see
Fig.~\ref{fig3}.  Unlike the behaviour on the one-body level,
$G_\parallel(x)$ under strong drive in model II does not resemble the
equilibrium mean-field result (obtained by calculating the
inhomogeneous magnetization around a fixed spin in the interface),
which displays significantly longer-ranged decay.

We studied coarse-grained interfacial properties via the local
position (height) of the interface $h(x)$, obtained either following
\cite{mullerInterface} or based on the (scaled) column magnetization:
$h(x)=(2m_{\rm b})^{-1}\sum_y \sigma(x,y)$; both methods give
consistent results. Fig.~3b displays results for the height-height
correlation function $C(x)=\langle h(x)h(0)\rangle$; this is directly
related to the spin-spin correlation function: $C(x)=(4m_{\rm
  b}^2)^{-1}\sum_{y,y'}G(x,y,y')$. The intercept, $C(0) = \langle
h(0)^2 \rangle$, provides an alternative measure of the (squared)
interfacial width; we find that $C(0)$ decreases under shear, very
similarly to the behaviour of $w^2$, exhibiting a very strong effect
for large values of $\omega$.  $C(x)$ decays faster with $x$ than in
equilibrium, in agreement with the behaviour of the spin-spin
correlation function. Hence from the scaling of $m(y)$ and
$G_\parallel(x)$, we expect $a_\perp^{-2} C(a_\parallel x) \approx
C_{\rm eq}(x)$, in correspondence with Weeks scaling in equilibrium
\cite{weeks}: $C_{\rm eq}(x) \approx w^2 {\cal C}(x/\xi_\parallel^{\rm
  eq})$, where $\cal C$ is the scaling function, and $w\sim
L_y$. Indeed both for strong boundary drive and weak bulk drive data
collapse is achieved using the {\em same} values for $a_\parallel$ and
$a_\perp$ as obtained above.

Comparing our findings for $C(x)$ to experimental results for the
height-height correlation function of a phase-separated
colloid-polymer mixture, Fig.~3 of Ref.~\cite{derks06prl}, reveals a
striking similarity in the reduction of the amplitude of the
correlation function under shear. This is consistent with the
observation of a reduction of the interfacial roughness in both
studies. However, Derks {\em et al.}  find an {\em increase} of the
lateral correlation length under shear (Fig.~4b of \cite{derks06prl}),
as obtained from fitting the correlation function of the equilibrium
capillary wave model to the steady state data. This finding is in
contrast to the behaviour of our models. Despite the differences in
the way the interface is localized (whether experimentally by gravity
or here by walls), the fundamental question arises as to which
features of interfaces under shear are universal. Clearly,
dimensionality is expected to play a major role.  The $3d$ Ising model
has a rough phase above the (finite) roughening transition
temperature. However, the roughness of the interface is established by
a very different mechanism than in $2d$, in that the interface throws
out spikes \cite{footnoteWhiskers}, which are expected to hinder fluid
flow to a much lesser degree than interface meandering in $2d$. This
should have important repercussions on the interfacial structure under
drive.

We thank D. Aarts, D. Derks, J. Eggers, R. Evans, A. Gambassi and
R. Zia for useful discussions and the EPSRC for support.

\clearpage
\end{document}